\documentclass{Rinton-P9x6}

\begin{document}

\title{Worldline Approach to the Casimir Effect}

\author{L.~Moyaerts, K.~Langfeld}

\address{Institut f\"ur Theoretische Physik, Universit\"at T\"ubingen\\
D-72076 T\"ubingen, Germany}

\author{H.~Gies}

\address{Institut f\"ur Theoretische Physik, Universit\"at Heidelberg\\
D-69120 Heidelberg, Germany}  

\maketitle

\abstracts{We present a new method to compute quantum energies in presence of 
a background field.  The method is based on the string-inspired worldline approach
to quantum field theory and its numerical realization with Monte-Carlo techniques. 
Our procedure is applied to the study of the Casimir force between rigid bodies induced
by a fluctuating real scalar field.  We test our method with the classic parallel-plate
configuration and study curvature effects quantitatively for the sphere-plate configuration.
The numerical estimates are compared with the ``proximity force approximation''.  
Sizable curvature effects are found for a distance-to-curvature-radius ratio of 
$a/R>0.02$.}

\section{Introduction}

This work is based on the formulation of the Casimir effect in the framework
of a renormalizable quantum field theory. The Casimir 
energy is accessed via the quantum effective action expressed in the 
string-inspired worldline approach to quantum field theory \cite{Gies:2003cv}.
Calculations are performed using a recently developed numerical method to 
compute quantum energies in presence of a background field \cite{Gies:2001tj}, which has
 already proved to be useful in the study of fermion-induced effects of magnetic interfaces
\cite{Langfeld:2002vy}.  In the next section, we derive the expression of the 
effective action on the worldline, define the Casimir energy and describe briefly 
the renormalization program.  Section~\ref{s:method} is devoted to the loop cloud 
method, our numerical implementation of worldline calculations. 
 Finally, we focus in section~\ref{s:results} 
on the study of Casimir forces between rigid bodies for the classic parallel-plate
configuration, which serves as test for the method, as well as for the experimentally 
relevant sphere-plate configuration. Curvature effects are investigated
and a comparison with the proximity force approximation is performed.

\section{Worldline formalism}

The embedding of the Casimir effect in quantum field theory, which is used in this 
approach and other contributions to this conference \cite{Graham:2002fw}, is 
an alternative to the standard approach of imposing {\it ab initio} boundary conditions on the fluctuating
field. By contrast, the quantum field theoretic approach permits a realistic modeling of the interactions
with the physical boundary and a careful study of the Casimir energy in 
the limiting case where boundary conditions are satisfied \cite{Jaffe:2003ji}.

We focus in this work on the Casimir effect induced by a fluctuating real scalar field.
The starting point is given by the field theoretic lagrangian
\begin{equation}
 \mathcal{L}=\frac{1}{2}\partial_\mu\phi\partial_\mu\phi+\frac{1}{2}m^2\phi^2
+\frac{1}{2}V(x)\phi^2. \label{eq:lag}
\end{equation}
Here we work in $D=d+1$ Euclidean spacetime dimensions, i.e. $d$ space dimensions. The interplay 
between the fluctuating field and the boundary matter is taken into account
by the interaction term $V(x)\phi^2$ where the potential $V(x)$ models the physical properties 
of the Casimir interface.  The complete unrenormalized quantum effective action for $V$ is
\begin{equation}
\Gamma[V]=\frac{1}{2}{\rm Tr\ ln}\frac{-\partial^2+m^2+V(x)}{-\partial^2+m^2}. \label{eq:effac}
\end{equation}
The worldline representation of $\Gamma[V]$ is obtained \cite{Schubert:2001he} by (i) introducing
 the propertime representation of the functional logarithm with UV regularization (e.g. a cutoff 
$\Lambda$ at the lower bound of the propertime integral), (ii) performing the trace {\it in 
configuration space}
 ${\rm Tr}[\hat{\mathcal{O}}] \to \int d^D x \langle x|\hat{\mathcal{O}}|x \rangle$
and (iii) interpreting the matrix element as a quantum mechanical transition amplitude 
expressed by a {\it Feynman path integral}.  The final expression reads
\begin{equation}
\Gamma[V]=-\frac{1}{2(4\pi)^{D/2}}\int_{1/\Lambda^2}^\infty
\frac{dT}{T^{D/2+1}}\  e^{-m^2 T}\!\! \int d^D x_{\rm CM} 
\left[\left\langle W_V[T,x(\tau)] \right\rangle_x-1\right], \label{eq:wlinerep}
\end{equation}
where the expectation value is defined by 
\begin{equation}
\left\langle W_V[T,x(\tau)] \right\rangle_x 
:=\frac{\int\limits_{x(0)=x(T)}\mathcal{D} x(\tau)\
 e^{-\int_0^T d\tau V(x_{\rm CM}+x(\tau))}
e^{-\int_0^T d\tau \frac{\dot{x}^2}{4}}}
{\int\limits_{x(0)=x(T)}\mathcal{D} x(\tau) \ e^{-\int_0^T d\tau \frac{\dot{x}^2}{4}}}. 
\label{eq:expvalue}
\end{equation}
This expression is the central piece of the worldline approach to quantum field theory. 
From Eq.(\ref{eq:wlinerep}), we see that the computation of the effective action corresponding 
to the theory (\ref{eq:lag}) with potential $V$ is based on the average of the holonomy factor
 $W_V[T,x(\tau)]$ over all {\it worldlines}  $x(\tau)$ which are parametrized by the propertime 
$\tau\in[0,T]$ and closed, $x(0)=x(T)$, due to the tracing operation in Eq.(\ref{eq:effac}).
 In Eq.(\ref{eq:wlinerep}), all worldline loops have a common center of mass $x_{\rm CM}$, 
which implies $\int_0^T d\tau x_\mu(\tau)=0$. 

In this work, we concentrate exclusively on static Casimir configurations, i.e. 
the modeling potential $$ V(x)=V({\bf x}) $$
is time independent. In this case, the propertime integrand itself does not depend
 on time and the time integration can be carry out trivially, $\int dx_{\rm CM}^0=L_{x_0}$,
 where $L_{x_0}$ denotes the ``volume'' in time direction.  We define the (unrenormalized)
 Casimir energy as 
\begin{equation}
\mathcal{E}_V=\Gamma[V]/L_{x_0}.\label{eq:energy}
\end{equation}

The effective action as given in 
Eq.(\ref{eq:wlinerep}) is an alternative formulation of the quantum field theory defined
from (\ref{eq:lag}).  For this reason, the divergence structure of the unrenormalized 
effective action on the worldline can be expressed in the language of conventional 
Feynman diagrams.  This is done by expanding the propertime integrand in Eq.(\ref{eq:wlinerep})
 for small values of $T$: 
\begin{equation}
\int d^Dx\, \langle W_V[T,x(\tau)]-1\rangle_x
= - T\int d^Dx\, V(x)+\frac{T^2}{2} \int d^Dx\, V(x)^2 +{\cal O}(T^3), \label{eq:ren1}
\end{equation}
which should be read together with the propertime factor $1/T^{D/2+1}$ of Eq.(\ref{eq:wlinerep}). 
For $D>2$, the first term 
is divergent and can always be eliminated by using the ``no-tadpole'' renormalization scheme. 
 The second term is divergent for $D\geq 4$ and contributes to the one-loop
diagram containing two insertions of the potential.  Standard renormalization provides
us with a counter term $\sim \int dx V^2$ subject to a physically chosen renormalization 
condition such that the divergence arising from the $T^2$ is canceled and which fixes
the physical value of the renormalized operator $\sim V^2$.

This procedure permits to handle these divergences which have a clear field theoretic 
origin.  Moreover further divergences may arise from the {\it modeling} 
of the Casimir configuration, i.e. from the potential $V$ itself.  
These divergences occur when the potential is tuned to take some limiting form 
$V\to V_{\rm cr}$ correspondingly to some convenient assumptions (ideal cases), 
$ \Gamma[V\to V_{\rm cr}]\to\infty $,
and may not be removed in a physically meaningful way.  The relevant question then is  
as to whether physical observables are affected by these divergences or not.  If
not, the problem can possibly be bypassed.

\section{Loop Cloud Method} \label{s:method}
As it is clear from Eq.(\ref{eq:wlinerep}), the main task to be performed is the average 
(\ref{eq:expvalue}) over the ensemble 
of all closed worldlines $x(\tau)$ centered upon a given point $x_{\rm CM}$.  The idea of the 
loop cloud method is to perform this expectation value by generating the loop 
ensemble in a numerical way \cite{Gies:2001tj}. It is evident that we can generate
 neither the whole loop ensemble, which is infinite, nor an arbitrarily big amount of loops,
 for clear reasons of computational time.  We will adopt a strategy which is widely used in
 statistical physics, namely the selection among the ensemble of all loops of the relevant
 configurations contributing to the loop average.  These are generated
 according to the Gau\ss ian loop distribution 
$p_{\rm loop}[T,x(\tau)]\simeq e^{-\int_0^T d\tau \frac{\dot{x}^2}{4}}$.
Generating an ensemble of $N_{\rm \,lp}$ such loops $\{x(\tau)|_i,\ i=1,\dots,N_{\rm lp}\}$,
 the expectation value (\ref{eq:expvalue}) is approximated by 
\begin{equation}
 \langle W_V[T,x(\tau)]\rangle_x\simeq \frac{1}{N_{\rm lp}}
\sum_{i=1}^{N_{\rm lp}} W_V[T,x(\tau)|_i]. \label{eq:mcarlo}
\end{equation}
The numerial burden is moreover dramatically reduced if we rescale the loops $x_\mu(\tau)$  
in such a way that the distribution $p_{\rm \,loop}$ becomes $T$-independent. Let us  consider
 the following rescaling transformations: 
$$ \tau\in[0,T]\to t:=\frac{\tau}{T}\in[0,1],\ \ \ x(\tau)\to y(t):=\frac{x(tT)}{\sqrt{T}}.$$
The propertime integral appearing in the loop distribution becomes 
$$ \int_0^T \!\!d\tau \ \dot{x}^2(\tau) \to \int_0^1\!\! dt \ \dot{y}^2(t) $$
where the $T$-dependence has indeed disappeared.  The numerical task is then reduced
 to the all-at-once generation \cite{Gies:2003cv} of an ensemble of $y$-loops, which are
 called {\it unit loops}.
  The evaluation of the expectation value (\ref{eq:mcarlo}) for a given value of $T$ consists
 then simply in the rescaling of all members of the unit loop ensemble 
$ x(\tau)=\sqrt{T}y(\tau/T) $
before the sum (\ref{eq:mcarlo}) is performed.
It is also obvious that the concept of continuous loops has to be abandoned.  In the simulations
 a loop is defined by a collection of $N_{\rm ppl}$ points, corresponding to the discretization 
of the propertime interval $[0,T]$:
$$ x(\tau)\to \left\{x_0=x(\tau_0),x_1=x(\tau_1),\ \dots\ ,x_{N_{\rm ppl}-1}
=x(\tau_{N_{\rm ppl}-1}),x_{N_{\rm ppl}}=x_0\right\}. 
$$
Let us point out that this discretization is not to be mistaken for a spacetime discretization. 
Every point $x_i$ can move freely in continuous spacetime. 

 The loop cloud method is developed independently of the background potential. This implies in
particular that the Casimir effect can be studied numerically for arbitrary geometries and independently
of the degree of symmetry of the physical boundaries.
This permits us to go beyond the 
geometry restrictions required by the 'proximity force approximation' (PFA) \cite{PFA1,PFA2}, 
which is the standard analytical approach to non trivial geometric configurations.
Beyond this, the method does not require perfect geometries and perfect surfaces but can, for instance,
facilitate the study of corrugated surfaces \cite{Emig:2002xz}.

\section{Casimir forces between rigid bodies} \label{s:results}

In this work, we model the physical boundaries by a potential of the form 
$$ V({\bf x})=\lambda f({\bf x}),$$
where $f({\bf x})$ describes the geometry of the Casimir configuration and 
$\lambda>0$ is the strength of the field-matter interaction.
It can roughly be viewed as a plasma 
frequency of the boundary matter: for fluctuations with frequency $\omega\gg\lambda$, 
the Casimir boundaries become transparent. Two types of limits can be taken regarding this
parametrization: 
\begin{eqnarray}
f({\bf x})&\to&\int_\Sigma d\sigma\  \delta^d({\bf x}-{\bf x}_\sigma)
\  {\rm (sharp\ potential\  limit)}
, \label{eq:sharp}\\
\lambda&\to&\infty\hspace{2.35cm} {\rm (strong\ coupling\  limit)}. \label{eq:dirich}
\end{eqnarray}
Here $\Sigma$ represents the geometry of the Casimir configuration and denotes a $d-1$ 
dimensional surface, generally disconnected (e.g. two separated plates, $\Sigma=S_1+S_2$). 
 The first limit idealizes the plate to be infinitely thin.  The second limit imposes 
the {\it Dirichlet boundary condition}, implying that all modes have to vanish on $\Sigma$.
Together they correspond to the case of {\it ab initio} boundary conditions of the 
standard approach. 
 These limits are unfortunately spoiled by the occurence of a divergence in the 
Casimir energy which cannot be removed by the standard field theoretic renormalization 
procedure.  
However, this problem can be bypassed if we consider instead the Casimir {\it force} 
acting on the rigid bodies.  The force is defined by 
$$ F_{\rm Cas}(a)=-\frac{\partial}{\partial a}E_{\rm Cas}(a) $$
where $a$ denotes the distance between the boundaries. This definition gives us a 
certain freedom of the choice of the {\it Casimir interaction energy} 
$E_{\rm Cas}(a)$.  For a Casimir configuration of two
 rigid bodies whose surfaces are denoted by $S_1$ and $S_2$, we define the interaction 
energy as  
\begin{equation}
 E_{\rm Cas}(a)=\mathcal{E}_{V_{1+2}}(a)-\mathcal{E}_{V_1}-\mathcal{E}_{V_2} 
\label{cef:intenergy}
\end{equation}
where  $\mathcal{E}_{V_{1+2}}$, $\mathcal{E}_{V_1}$ and $\mathcal{E}_{V_2}$ are the energies 
(\ref{eq:energy}) with the potentials $V_{1+2}$, $V_1$ and  $V_2$ given in the limiting form 
(\ref{eq:sharp}) where $\Sigma=S_1\cup S_2$, $\Sigma=S_1$ and $\Sigma=S_2$, respectively. 
The first term depends on the distance $a$ and 
contributes to the Casimir force.  Since the remaining terms do not depend on $a$, 
the way there are chosen has no influence on the Casimir force.  In the present case, 
they are chosen in such a way that $E_{\rm Cas}(a)$ is rendered finite by substracting
 each Casimir energy of the single bodies.

\begin{figure}[t]
\epsfxsize=13pc
\epsfbox{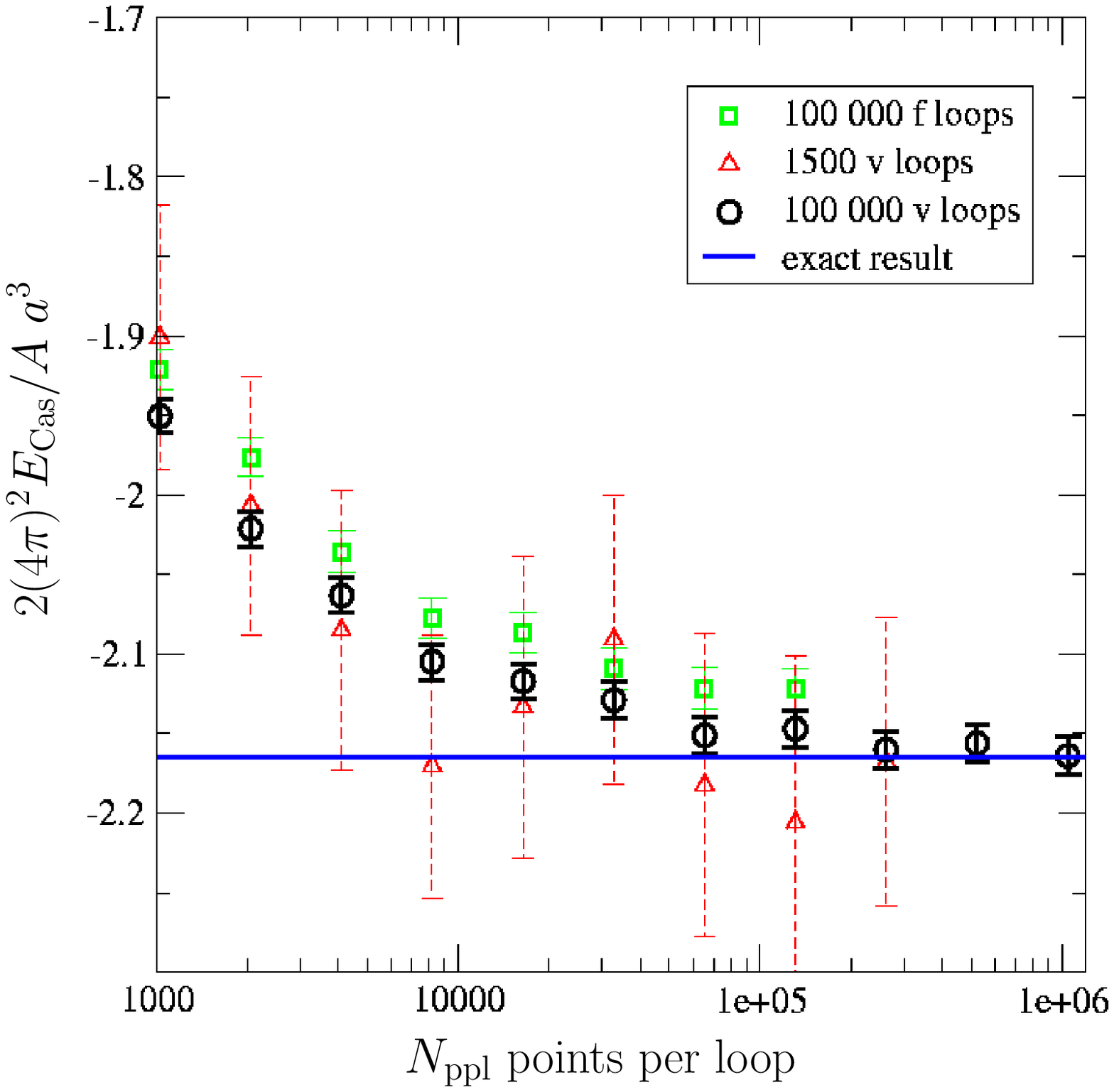} 
\epsfxsize=14pc
\epsfbox{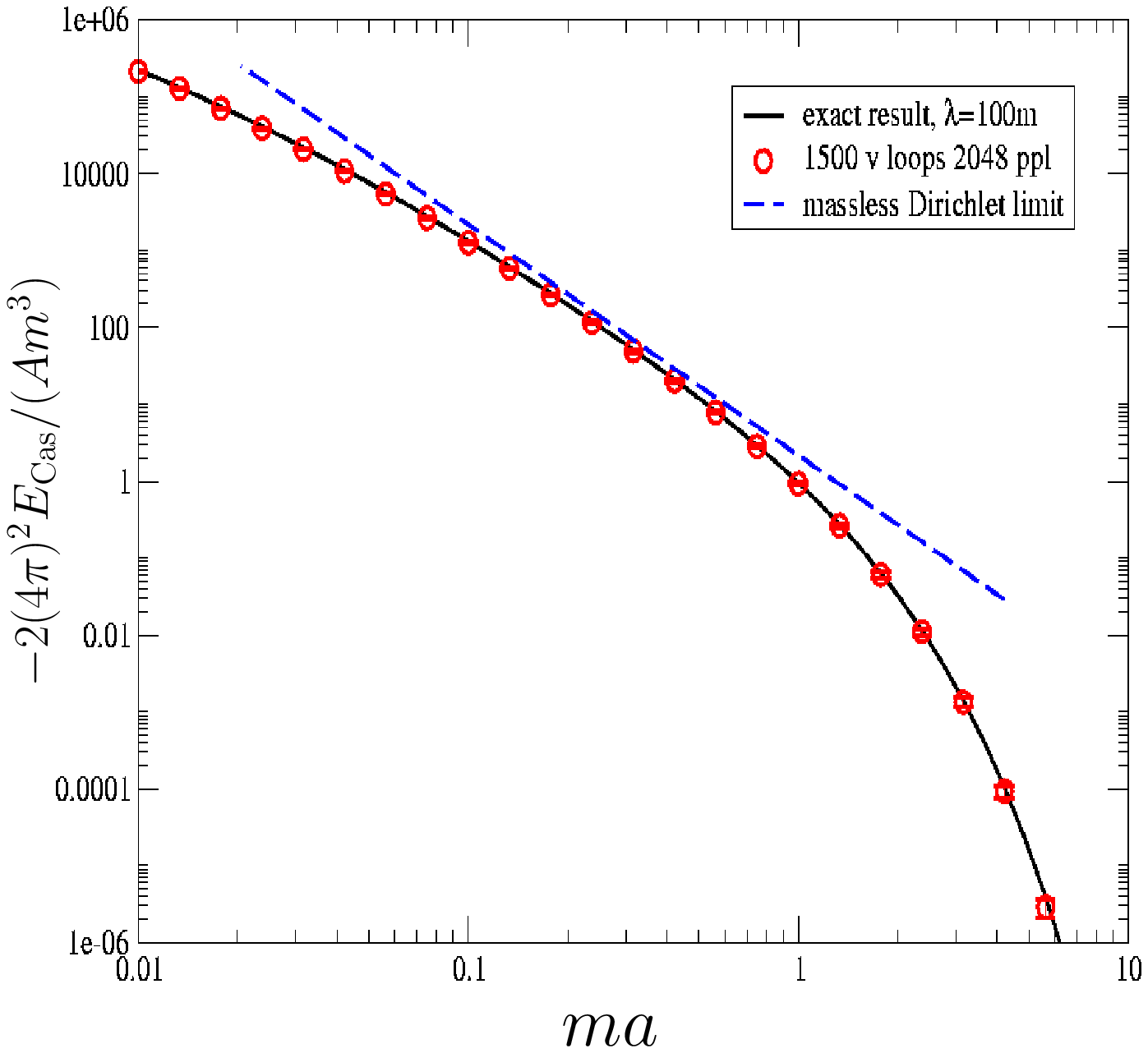}
\caption{Numerical estimate of the interaction Casimir energy per unit area of the
parallel-plate configuration for various loop ensembles as a function
of the number of points per loop $N_{\rm ppl}$ (left panel).  Numerical estimates vs. 
analytical results as a function of the dimensionless quantity $ma$, $\lambda=100 m$ (right panel).}
\label{fig:test}
\end{figure}

\subsection{A benchmark test: the parallel plates revisited}
Casimir forces can be analytically computed only for a small number of rigid-body 
geometries among which there is the parallel-plate configuration. Comparing our numerical
 estimates 
with the analytically known result \cite{Bordag:1992cm}, the agreement 
is very satisfying in all regimes of scalar mass $m$, coupling $\lambda$ and distance 
$a$ , see Fig.\ref{fig:test}, right panel.  Moreover, we have found
 that our algorithm is scalable: if higher precision is 
required, only the parameters of the loop ensemble, points per loop $N_{\rm ppl}$ which 
controls the systematic error induced by the loop discretization and number of loops
$N_{\rm lp}$ which controls the approximation of the loop average (\ref{eq:expvalue})
by the sum (\ref{eq:mcarlo}), have to be adjusted, as seen in Fig.\ref{fig:test}, left
panel.

\subsection{Beyond the PFA: the sphere-plate configuration}
\begin{figure}[t]
\epsfxsize=17pc 
\centerline{\epsfbox{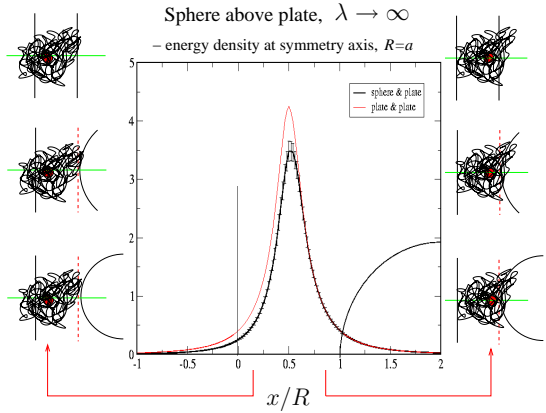}} 
\caption{Interaction Casimir energy density
along the symmetry axis ($x$ axis) for the sphere-plate configuration
in comparison to the parallel-plate case. Close to the sphere, the
worldline loops do not ``see'' the curvature; but at larger distances,
curvature effects enter the energy density. For illustration, the
sphere-plate geometry is also sketched (thin black lines).}
\label{fig:loops}
\end{figure}

We study the experimentally relevant sphere-plate configuration and confront our 
numerical estimates with the results provided by the proximity force approximation (PFA).
Due to the assumptions on which the PFA is based, i.e. dropping of non-local curvature
 effects,
 this approach is expected to give reasonable
results only if the typical curvature radii of the surfaces
elements is large compared to the element distance.  
 Beside its physical interest, the sphere-plate
configuration with its relatively simple geometry provides us with a beautiful illustration 
of the loop cloud machinery. As illustrated in Fig.\ref{fig:loops}, the main feature of the 
sphere-plate physics, i.e. 
curvature effects, can be understood qualitatively simply by thinking in
 terms of loops.
On the one hand, the energy density at a point $x_0$ is obtained by performing averages over
 loop clouds centered on $x_0$.  On the other hand, it can be shown that the energy density 
is proportional to the number 
of loops piercing both physical boundaries.  With these two statements only, 
Fig.(\ref{fig:loops}) can be qualitatively understood.   The effect of the increasing
 curvature
 of one of 
the boundaries is the decrease of contributing loops near the plate, implying the decrease
 of the Casimir energy in this region.

\begin{figure}[t]
\epsfxsize=13pc 
\epsfbox{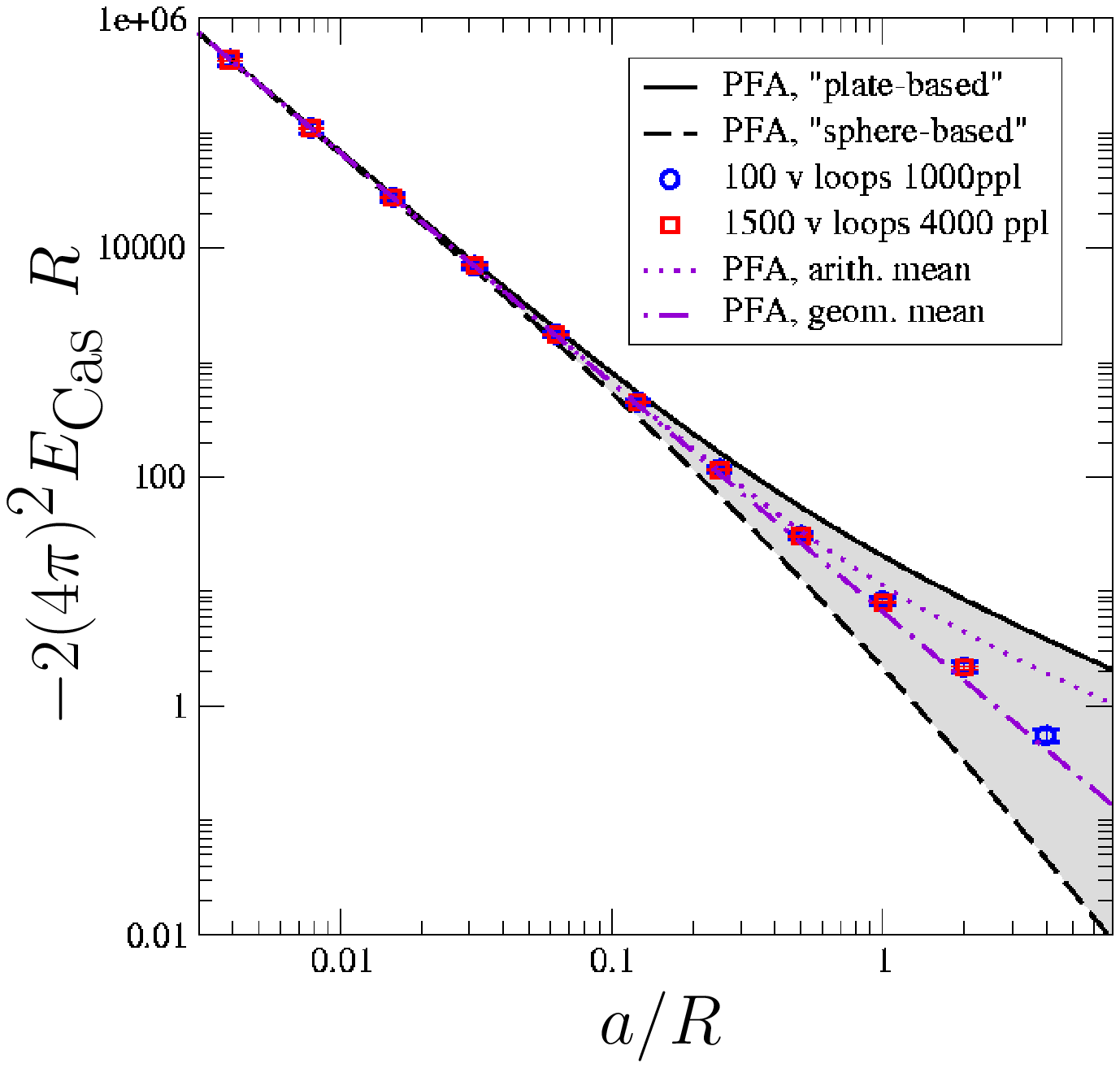} 
\epsfxsize=14pc
\epsfbox{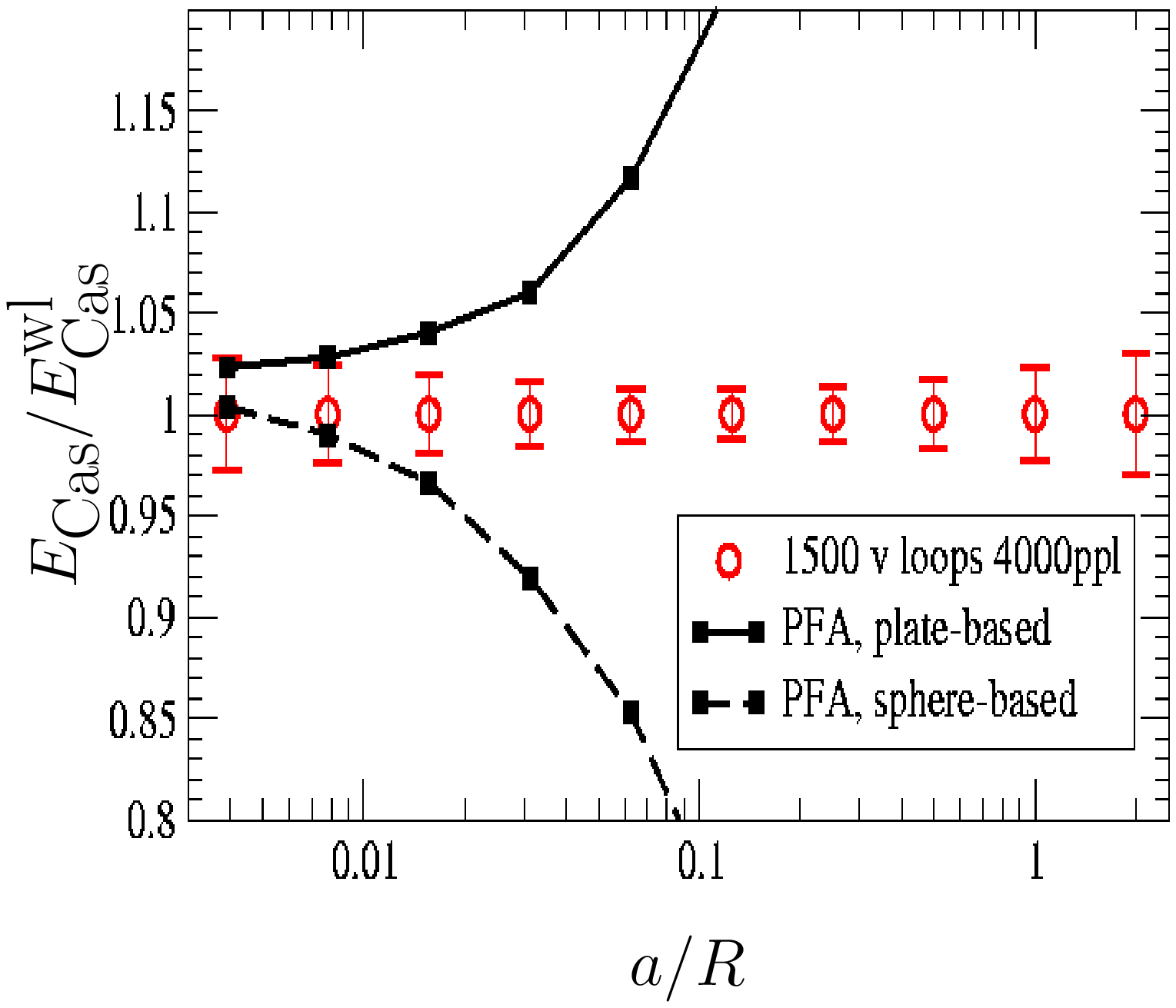}
\caption{{\bf Left panel} Logarithmic plot of the interaction
Casimir energy for the sphere-plate configuration. For small
separations/large spheres, $a/R<0.02$, the proximity force
approximation (PFA) approximates the numerical estimate well; but for larger
$a/R$, curvature effects are not properly taken into account. The PFA
becomes ambiguous for larger $a/R$, owing to possible different choices
of the integration domain in the PFA. A geometric mean
(dotted-dashed line) of the limiting cases 
shows reasonable agreement with the numerical result. {\bf Right panel} Interaction Casimir energies
normalized to the numerical result . For $a/R>0.02$, the fluctuation-induced
curvature effects occur at the percent level.}
\label{fig:sap}
\end{figure}

Let us finally consider the complete interaction Casimir energy for the
sphere-plate configuration as a function of the sphere-plate distance
$a$ (we express all dimensionful quantities as a function of the
sphere radius $R$). In Fig.~\ref{fig:sap}, we plot our numerical
results in the range $a/R\simeq {\cal O}(0.001\dots 10)$. Since the
energy varies over a wide range of scales, already small loop
ensembles with rather large errors suffice for a satisfactory
estimate (the error bars of an ensemble of 1500 loops with 4000 ppl
cannot be resolved in Fig.~\ref{fig:sap}). 
Let us compare our numerical estimate with the proximity force
approximation with radial distance measurements: using the plate surface as the integration 
domain
for the PFA, we obtain the solid line in
Fig.~\ref{fig:sap} (PFA, plate-based), corresponding to a
``no-curvature'' approximation. As expected, the PFA approximation
agrees with our numerical result for small distances (large sphere
radius). Sizable deviations from the PFA approximation of the order of
a few percent occur for $a/R\simeq0.02$ and larger. Here, the
curvature-neglecting approximations are clearly no longer valid. This
can be read off from the right panel, where the resulting
interaction energies are normalized to the numerical result.

\section{Conclusion}
In this work, we have developed a new numerical method to compute 
quantum energies in presence of a background field.  The algorithm 
is constructed independently of the background, which opens a 
wide range of applicability of the procedure.  In the context 
of the Casimir effect, arbitrary geometric configurations can be 
considered.  We focused in this work on the Casimir forces induced
on rigid bodies by the fluctuations of a real scalar field.  The method 
was tested by comparing our numerics with the analytical results 
in the classic parallel-plates configuration and confronted with 
the results provided by the proximity force approximation in the 
sphere-plate configuration. We have been able to study the usually
neglected nonlocal curvature effects which become sizable for a 
distance-to-curvature-radius ratio of $a/R> 0.02$. Some generalizations
to more realistic configurations are straightforward: finite temperature
effects, surface roughness and corrugation. 
Let us finally note that the implementation of worldline
numerics for the Casimir effect due to a fluctuating electromagnetic field
has still to be worked out. Here the new ingredient is to find an
efficient field theoretic formulation of the interaction of the
electromagnetic field with the Casimir boundaries.

\section*{Acknowledgments}

L.M. is grateful to the organizers of the QFEXT03 meeting in Norman for the
stimulating atmosphere of the conference and  acknowledges financial support 
by the Deutsche Forschungsgemeinschaft under contract GRK683.
H.G. is supported by the Deutsche Forschungsgemeinschaft
under contract Gi 328/1-2.

\end{document}